\documentclass[letterpaper,prl,aps,twocolumn,showpacs,superscriptaddress]{revtex4-1}
\usepackage{bm,color}
\usepackage{mathptmx}
\usepackage[T1]{fontenc}
\usepackage[utf8]{inputenc}
\usepackage{graphicx} 
\usepackage{amsmath} 
\usepackage{amsfonts} 

	\newcommand{\com}{\quad\text{,}}
	\newcommand{\dt}{\quad\text{.}}
	\newcommand\numberthis{\addtocounter{equation}{1}\tag{\theequation}}
	\newcommand{\ave}[1]{\langle #1 \rangle}
	\newcommand{\vb}[1]{\left( #1 \right)}
	\newcommand{\vsb}[1]{\left[ #1 \right]}
	
	\newcommand{\mbf}[1]{\mathbf{#1}}
	\newcommand{\bbm}{\begin{pmatrix}}
	\newcommand{\ebm}{\end{pmatrix}}	
	\newcommand{\ve}[1]{\hspace{0.5pt}\mbf{#1}}

	\usepackage[normalem]{ulem} 
	\definecolor{mgreen}{RGB}{1,123,0}

\begin{document}

\title{Observation of topological Bloch-state defects and their merging transition}

\author{Matthias Tarnowski}
\thanks{These authors have contributed equally to this work}
\affiliation{Institut für Laserphysik, Universität Hamburg, 22761 Hamburg, Germany}
\affiliation{The Hamburg Centre for Ultrafast Imaging, 22761 Hamburg, Germany}
\author{Marlon Nuske}
\thanks{These authors have contributed equally to this work}
\affiliation{Zentrum für Optische Quantentechnologien, Universität Hamburg, 22761 Hamburg, Germany}
\author{Nick Fläschner}
\affiliation{Institut für Laserphysik, Universität Hamburg, 22761 Hamburg, Germany}
\affiliation{The Hamburg Centre for Ultrafast Imaging, 22761 Hamburg, Germany}
\author{Benno Rem}
\affiliation{Institut für Laserphysik, Universität Hamburg, 22761 Hamburg, Germany}
\affiliation{The Hamburg Centre for Ultrafast Imaging, 22761 Hamburg, Germany}
\author{Dominik Vogel}
\affiliation{Institut für Laserphysik, Universität Hamburg, 22761 Hamburg, Germany}
\author{Lukas Freystatzky}
\affiliation{Zentrum für Optische Quantentechnologien, Universität Hamburg, 22761 Hamburg, Germany}
\author{Klaus Sengstock}
\email{klaus.sengstock@physnet.uni-hamburg.de}
\affiliation{Institut für Laserphysik, Universität Hamburg, 22761 Hamburg, Germany}
\affiliation{The Hamburg Centre for Ultrafast Imaging, 22761 Hamburg, Germany}
\affiliation{Zentrum für Optische Quantentechnologien, Universität Hamburg, 22761 Hamburg, Germany}
\author{Ludwig Mathey}
\affiliation{Institut für Laserphysik, Universität Hamburg, 22761 Hamburg, Germany}
\affiliation{The Hamburg Centre for Ultrafast Imaging, 22761 Hamburg, Germany}
\affiliation{Zentrum für Optische Quantentechnologien, Universität Hamburg, 22761 Hamburg, Germany}
\author{Christof Weitenberg}
\affiliation{Institut für Laserphysik, Universität Hamburg, 22761 Hamburg, Germany}
\affiliation{The Hamburg Centre for Ultrafast Imaging, 22761 Hamburg, Germany}

\date{\today}

\pacs{67.85.-d, 67.85.Lm}

\begin{abstract}
Topological defects in Bloch bands, such as Dirac points in graphene, and their resulting Berry phases play an important role for the electronic dynamics in solid state crystals. Such defects can arise in systems with a two-atomic basis due to the momentum-dependent coupling of the two sublattice states, which gives rise to a pseudo-spin texture. The topological defects appear as vortices in the azimuthal phase of this pseudo-spin texture. Here, we demonstrate a complete measurement of the azimuthal phase in a hexagonal optical lattice employing a versatile method based on time-of-flight imaging after off-resonant lattice modulation. Furthermore we map out the merging transition of the two Dirac points induced by beam imbalance. Our work paves the way to accessing geometric properties in optical lattices also with spin-orbit coupling and interactions.
\end{abstract}

\maketitle

\begin{figure}[tb]
		\includegraphics[width=\linewidth]{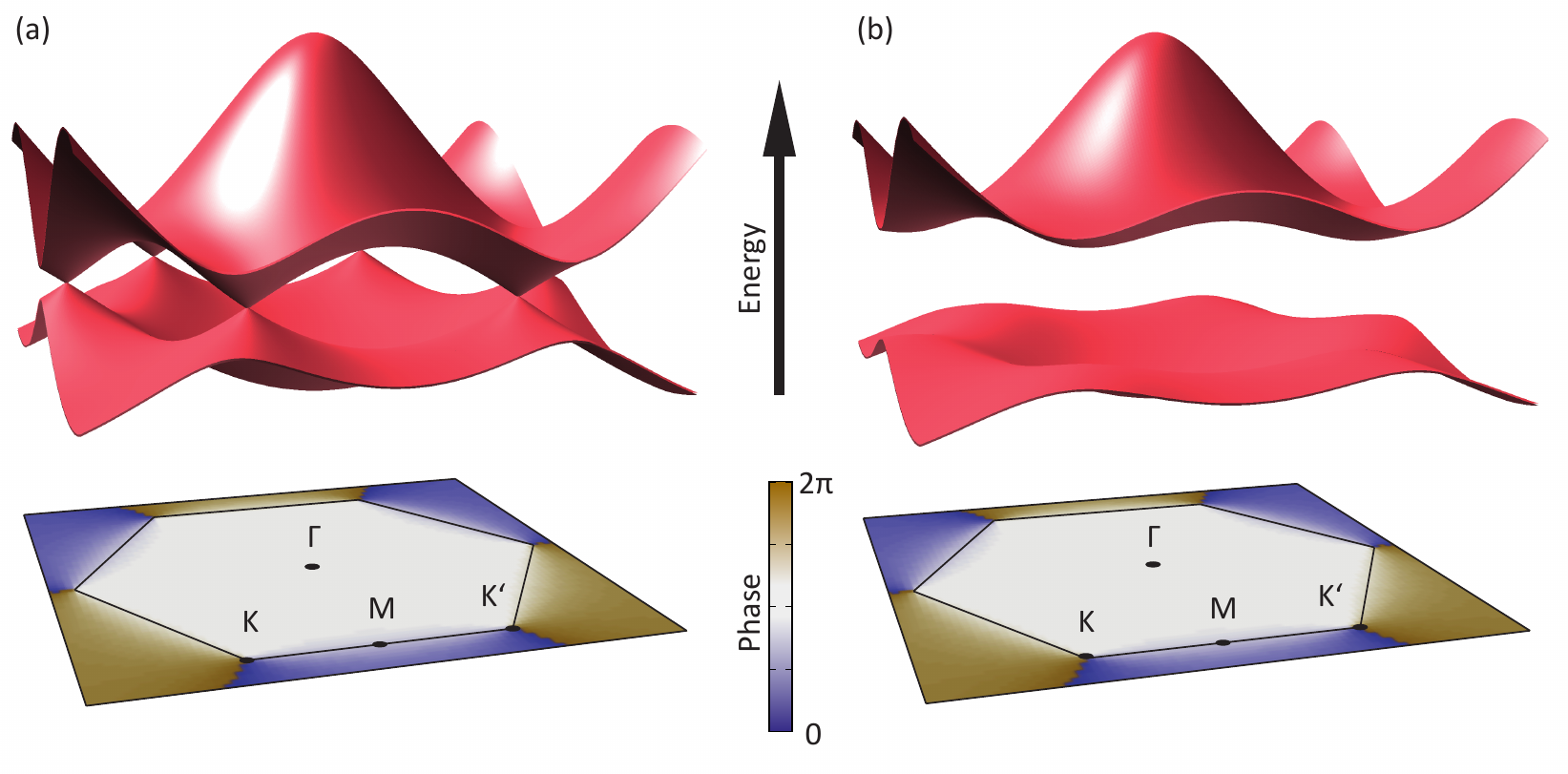}
		\caption{Topological defects in Bloch bands. a) The dispersion relation in momentum space of the two lowest bands of a graphene lattice. The bands touch linearly at the two Dirac points, i.e. the K and $\rm K'$ points, at the edge of the Brillouin zone (hexagon). The corresponding eigenstates form a pseudo-spin texture in momentum space with the azimuthal phase $\phi_{\ve k}$ (projection plane below). The vortices of the azimuthal phase show that the Dirac points are topological defects in the pseudo-spin texture. b) When the inversion symmetry is broken by an energy offset between the sublattices, the Dirac points open and become massive. The topological defects remain unchanged.}\label{fig:1_topological_defects}
		\end{figure}

	The motion of a particle in a crystal is not only affected by the band dispersion,  but also by the geometry of the Bloch states. 
	This geometrical property of the states was famously pointed out by Thouless and Berry \cite{thouless_quantized_1982,berry_quantal_1984}, and constitutes a fundamental feature of crystalline structures in their tremendously diverse forms. Furthermore, this geometric structure, which is captured by the Berry curvature, can have singular features, which have a topological nature. The paradigmatic example of such topological defects are the Dirac points in graphene, shown in Fig.~\ref{fig:1_topological_defects}. In a two-band model, the geometry of the eigenstates can be visualized as a pseudo-spin 1/2 texture in momentum space. The topological defects are quantized vortices in the azimuthal phase, that are located at the Dirac points. These features are indeed responsible for  the special electronic transport properties of graphene, see \cite{xiao_berry_2010}.
 	Beyond graphene and its properties, topological defects also control numerous other intriguing phenomena in solid state physics, such as the integer quantum Hall effect \cite{klitzing_new_1980} or topological insulators \cite{hasan_colloquium_2010}.

	In recent years, ultracold atoms in optical lattices have emerged as a versatile model system with tunable topological properties \cite{goldman_topological_2016}. Besides the measurement of the band gap at topological defects \cite{tarruell_creating_2012,weinberg_breaking_2016}, they also offer detection tools, which access the eigenstates similar to ARPES measurements in graphene \cite{gierz_graphene_2012}. For example, accelerating wave packets through the lattice gives access to diverse geometric phases via interferometry \cite{atala_direct_2013,duca_2014,li_bloch_2016} or to global topological invariants such as the Chern number via differential drift measurements \cite{price_mapping_2012,jotzu_experimental_2014,aidelsburger_measuring_2015}. Finally working with a filled lowest band, a projection onto flat bands yields a full state tomography, from which topological defects and global topology can be determined \cite{hauke_tomography_2014,flaschner_experimental_2016}. These methods are, however, either inefficient for covering the full Brillouin zone, cannot resolve the position of the defects, or only work in specific systems such as Floquet systems.

	\begin{figure}[tb]
		\includegraphics[width=\linewidth]{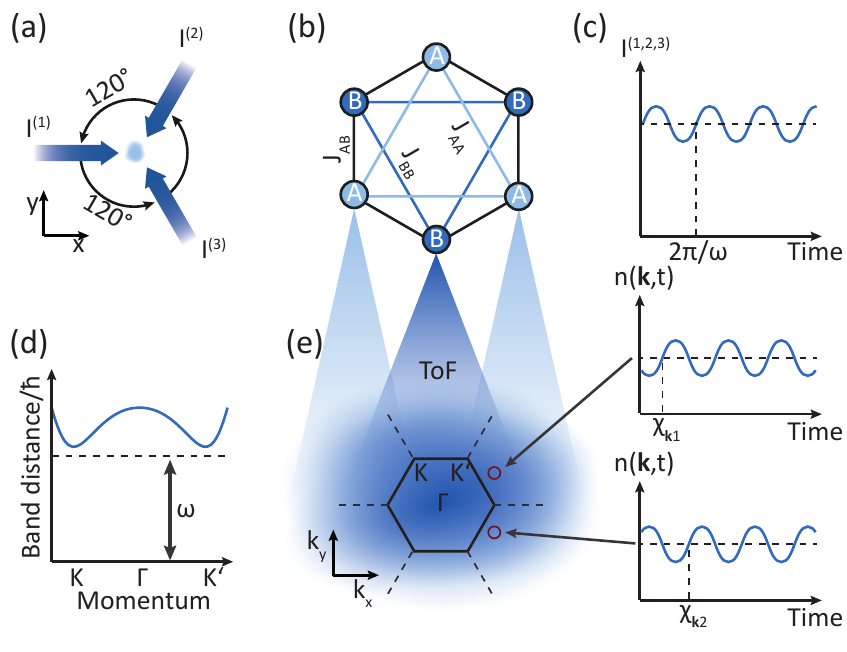}
		\caption{Detection of the azimuthal phase structure via lattice modulation. a) The interference pattern of three laser beams with intensities $I_1$, $I_2$ and $I_3$ forms a hexagonal lattice with lattice sites A and B on two triangular sublattices. b) In a tight-binding model, it can be described by tunneling between nearest and next-nearest neighbors with amplitude $J_{AB}$ and $J_{AA}, J_{BB}$, respectively, and by a sublattice energy offset $\Delta_{AB}$ (indicated by different colors), which is created by using appropriate polarizations of the laser beams \cite{flaschner_experimental_2016}. c) The laser beam intensities are modulated with a driving frequency $\omega$ to probe the temporal response of the momentum states. d) Sketch of the energy distance between the two lowest bands. The driving frequency is chosen red-detuned to the band gap. e) We measure the momentum space density $n({\mbf k},t)$ after ToF expansion for various modulation times $t$ and obtain an oscillation with distinct phase $\chi_{\ve k}$ for each momentum $\mbf k$. The measured phase $\chi_{\ve k}$ is directly related to the azimuthal phase $\phi_{\ve k}$ of the eigenstates.}\label{fig:2_detectionScheme}
	\end{figure}

	Here, we present a measurement of the topological defects of the Bloch states of a hexagonal optical lattice. We map out the azimuthal phase profile of the pseudo-spin texture in entire momentum space and identify the phase windings as topological defects. We introduce a new versatile method to measure this phase for all momentum states simultaneously, which also works for a completely filled first band. Off-resonant modulation of the laser beam intensities leads to a modulation of the momentum space density. We show that the phase of this modulation is related to the azimuthal phase.  As a compelling example, we realize the merging transition of two topological defects by varying the lattice beam intensities \cite{zhu_simulation_2007} and map out the transition by following their position.

	We consider a system of ultracold atoms in a hexagonal optical lattice \cite{becker_ultracold_2010,flaschner_experimental_2016} (see Fig.~\ref{fig:2_detectionScheme}(a)). The lattice consists of two triangular sublattices, labeled as A and B, with the annihilation operators for the momentum states ${\ve k}$ on the respective sublattices $c_{{\ve k},A}$ and $c_{{\ve k},B}$. Neglecting a momentum-dependent energy offset, the tight-binding Hamiltonian is given by $H_{\rm tb}=\sum_{\ve k} H_{\rm tb,{\ve k}}$, where
	\begin{align}
		H_{\rm tb,{\ve k}}&=\epsilon_{\ve k} \bbm c_{{\ve k},A}^\dag & c_{{\ve k},B}^\dag \ebm \bbm \cos(\theta_{\ve k}) & \sin(\theta_{\ve k})e^{-i\phi_{\ve k}} \\ \sin(\theta_{\ve k})e^{i\phi_{\ve k}} & -\cos(\theta_{\ve k}) \ebm \bbm c_{{\ve k},A} \\ c_{{\ve k},B}\ebm \label{eq:hamiltonianThetaPhi}
	\end{align}
	and describes a pseudo-spin with azimuthal phase $\phi_{\ve k}$ and polar angle $\theta_{\ve k}$ for each momentum. The band distance is $2\epsilon_{\ve k}$. The 
	eigenstates with annihilation operators 
	\begin{align} 
		\begin{split}
			c_{{\ve k},+}&=\cos(\theta_{\ve k}/2)c_{{\ve k},A}+\sin(\theta_{\ve k}/2)e^{-i\phi_{\ve k}}c_{{\ve k},B}\\
			c_{{\ve k},-}&=-\sin(\theta_{\ve k}/2)e^{i\phi_{\ve k}}c_{{\ve k},A}+\cos(\theta_{\ve k}/2)c_{{\ve k},B} \label{eq:lowerUpperBandOperators}
		\end{split}
	\end{align}
	describe the two lowest bands of the lattice. The dependence of $\theta_{\ve k}$, $\phi_{\ve k}$ and $\epsilon_{\ve k}$ on the tight-binding parameters $J_{AB}$, $J_{AA}$, $J_{BB}$, $\Delta_{AB}$ (see Fig.~\ref{fig:2_detectionScheme}) is given in \cite{supmat}. The vortices of $\phi_{\ve k}$ indicate the topological defects of the Bloch states.

	In order to access the phase $\phi_{\ve k}$, we employ a new method, which is based on the interference of the A and B sublattices in time-of-flight (ToF) expansion after off-resonant lattice modulation. The density distribution after ToF is up to a Wannier envelope given by
	\begin{align*}
		n(\mbf k)=n_{{\ve k},A}+n_{{\ve k},B} + (\ave{c_{{\ve k},A}^\dag c_{{\ve k},B}}+c.c.) \numberthis\label{effTofExp}\com
	\end{align*}
	where $n_{{\ve k},A}=\ave{c_{{\ve k},A}^\dag c_{{\ve k},A}}$ and $n_{{\ve k},B}=\ave{c_{{\ve k},B}^\dag c_{{\ve k},B}}$ are the occupations of the two sublattices. Rewriting it in terms of the eigenstates corresponding to the upper and lower bands yields
	\begin{align}
 		 n(\ve k)  &= A_{\mbf k, +} n_{\ve k, +} + A_{\mbf k, -} n_{\ve k, -} + (B_{\ve k} \langle c_{\ve k, +}^{\dagger} c_{\ve k, -}\rangle + c.c.)	\label{eq:nofkthetaphi}
 	\end{align}
 	where $n_{{\ve k},+}=\ave{c_{{\ve k},+}^\dag c_{{\ve k},+}}$ and $n_{{\ve k},-}=\ave{c_{{\ve k},-}^\dag c_{{\ve k},-}}$ are the occupations of the two bands. The prefactors $A_{\mbf k, \pm} = (1 \mp \cos \phi_{\ve k} \sin \theta_{\ve k})/2$ and $B_{\ve k}=(\cos(\theta_{\ve k})\cos(\phi_{\ve k})+i\sin(\phi_{\ve k}))/2$ contain information about the azimuthal phase $\phi_{\ve k}$, which however cannot be disentangled by a single measurement of the density. The crucial idea is to extract the phase from a measurement of the temporal response to a driving, which affects the terms differently. We employ off-resonant modulation of the laser beam intensities with a red-detuned driving frequency $\omega<\Delta_{AB}$ (see Fig.~\ref{fig:2_detectionScheme}(b)). For a discussion of other driving regimes see supplementary material \cite{supmat}.
 	
	For weak off-resonant driving, the band occupations remain largely unaffected, while the correlator $\ave{c_{{\ve k},+}^\dag c_{{\ve k},-}}$ becomes time dependent. The macroscopic occupation of the lowest band serves as a homodyne amplification of the very small induced occupation of the upper band, such that the correlator can be measured. To first order in perturbation theory and neglecting a fast oscillating part (see \cite{supmat}) the resulting density oscillation after ToF is:
	\begin{align}
		n(\mbf k,t)& \approx n_{{\rm eq},{\ve k}}- \delta n_{\ve k} \sin(\omega t +\chi_{\ve k}) \com \label{eq:ntof}
	\end{align}
	where $n_{{\rm eq},{\ve k}}$ is the ToF density without a lattice modulation and 
	\begin{align}
		\chi_{\ve k}&={\rm Arg}\vsb{\cos(\phi_{\ve k}) + i P_{\hspace{0.5pt}\ve k} \sin(\phi_{\ve k})}\dt\label{eq:chik}
	\end{align}
	The oscillation amplitude $\delta n_{\ve k}$ and the coefficient $P_{\ve k}$ are defined in \cite{supmat}.
	The measured phase $\chi_{\ve k}$ is closely related to the azimuthal phase $\phi_{\ve k}$. In particular, $\chi_{\ve k}$ smoothely depends on $\phi_{\ve k}$ and the position of their vortices is identical. Therefore the measurement of $\chi_{\ve k}$ gives a reliable determination of the position and winding number of the topological defects of the Bloch states. The distortion factor $P_{\ve k}$ is given by $P_{\ve k}\approx \omega/\Delta_{AB}$, and can be made close to one for near-resonant driving ($\omega\approx\Delta_{AB}$).

	\begin{figure}[htb]
		\includegraphics[width=\linewidth]{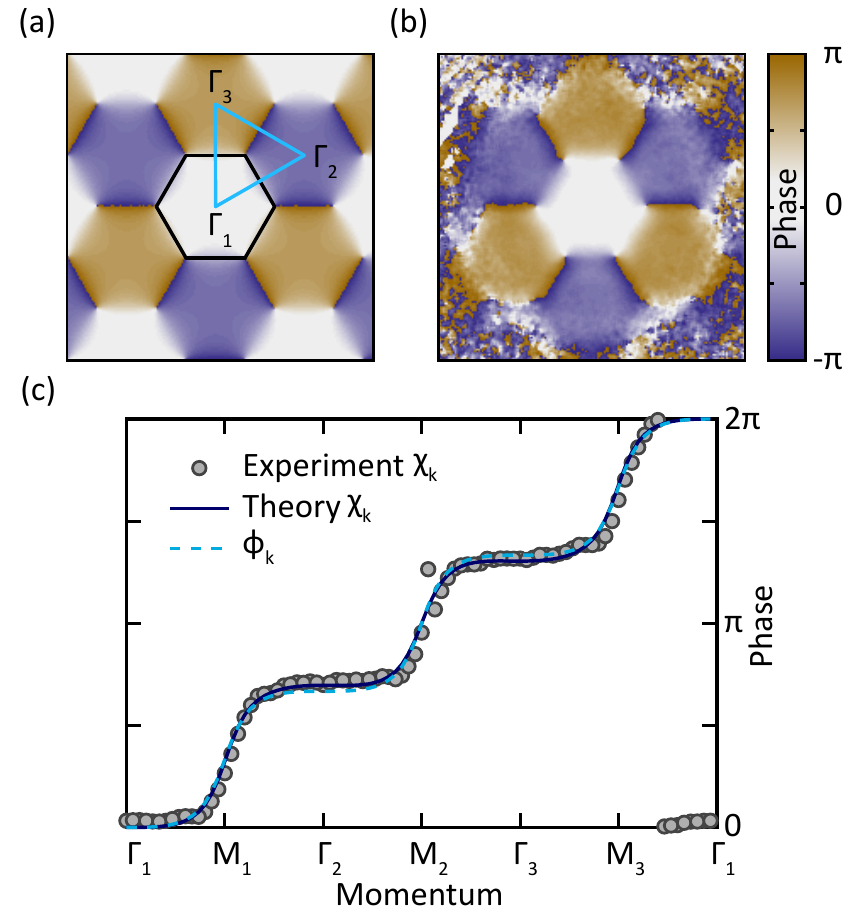}
		\caption{Measurement of the azimuthal phase profile. a) Expected azimuthal phase $\phi_{\ve k}$ in momentum space. The black hexagon marks the first Brillouin zone. b) Experimentally obtained phase $\chi_{\ve k}$. The position and winding number of the topological defects are correctly determined. c) Quantitative comparison of the phases along the high symmetry path indicated by the solid, blue lines in a). The data is averaged over the three equivalent paths of which only one is shown in a). The parameters are $\omega=2\pi\cdot 5500$ Hz, $J_{AB}=h\cdot 520$ Hz, $J_{AA}=h\cdot 99$ Hz, $J_{BB}\approx 0$ and $\Delta_{AB}=h\cdot 6056$ Hz (see \cite{supmat}). The laser beam intensities are modulated by $\pm 20\%$, leading to a modulation of $J_{AB}$ by $\pm 18\%$ and of $\Delta_{AB}$ by $\pm 22\%$. The experimentally measured $\chi_{\ve k}$ matches well with the numerical expectation, which we also find for other red-detuned driving frequencies. Furthermore, for the chosen parameters, the difference between $\chi_{\ve k}$ and $\phi_{\ve k}$ is experimentally indiscernable.}\label{fig:3_comparisonExperimentTheory}
	\end{figure}

	In Fig.~\ref{fig:3_comparisonExperimentTheory}b we show the full phase profile obtained with the method as described above. We start with a spin-polarized cloud with $10^5$ fermionic potassium atoms. We adiabatically ramp up a hexagonal optical lattice in a boron-nitride configuration with finite offset $\Delta_{AB}$, such that the lowest band is completely filled. We then suddenly apply a modulation of the laser beam intensities and measure the resulting ToF density distribution after variable modulation times. We extract $\chi_{\ve k}$ from a pixel-wise fit to the data and thus obtain the phase information with high resolution throughout momentum space. We compare the measurement to the prediction from perturbation theory for $\chi_{\ve k}$ and find very good agreement (Fig.~\ref{fig:3_comparisonExperimentTheory}). Due to the near-resonant choice of the driving frequency, the distortion between $\phi_{\ve k}$ and $\chi_{\ve k}$ is very small and we effectively measure $\phi_{\ve k}$. This phase profile illustrates the threefold symmetry of the eigenstates, which stems from the high symmetry of the lattice in real space \cite{lim_2015,flaschner_experimental_2016,li_bloch_2016}.

	\begin{figure}
		\includegraphics[width=\linewidth]{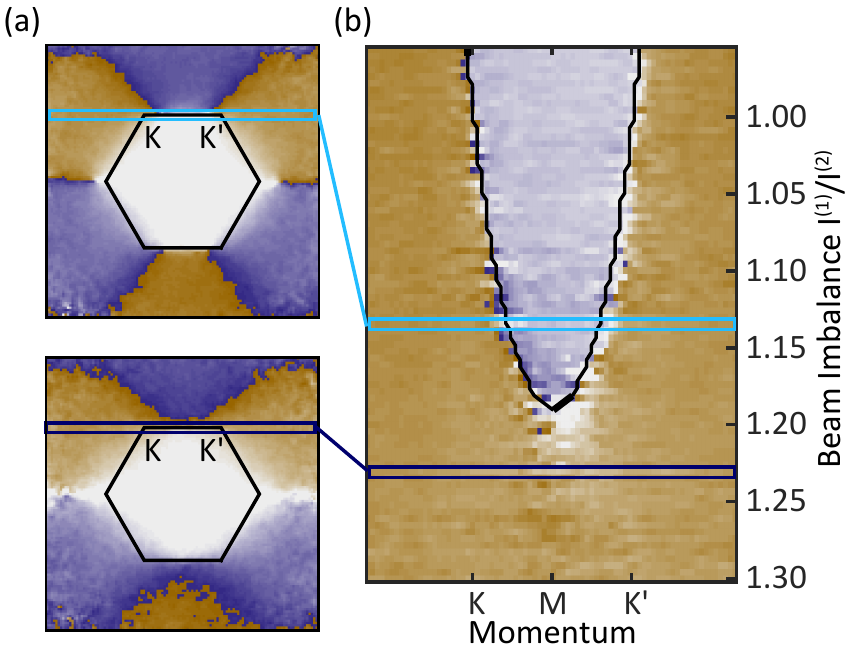}
		\caption{Mapping out the merging transition of topological defects. a) Phase profile for two different beam imbalances ($I^{(1)}/I^{(2)}=1.14$ and $I^{(1)}/I^{(2)}=1.23$). The hexagon marks the Brillouin zone. The blue boxes show the cut along $k_x$, which is displayed in b). b) movement of the topological defects (visible as $\pi$ phase jump in the cut) as a function of beam imbalance. The defects move along the K-M-$\rm K'$ path and merge at M point for a critical beam imbalance $I^{(1)}/I^{(2)}$ = 1.18. Lattice parameters as in Fig.~\ref{fig:3_comparisonExperimentTheory}, driving frequency is $\omega = 2\pi\cdot 4500$ Hz. The black line shows the positions of the defects as obtained from the local minima in a band structure calculation. 
	}\label{fig:4_mergingTransition}
	\end{figure}

In a second set of experiments, we use the same method to fully map out the merging transition of the topological defects. This can be achieved by changing the relative beam intensities of the optical lattice \cite{zhu_simulation_2007} or via lattice shaking \cite{koghee_2012}, which is equivalent to applying strain in a solid state material such as graphene \cite{amorim_novel_2016}. While strain is limited in a solid state system and in particular graphene doesn't hold the stress needed to reach the merging transition \cite{montambaux_merging_2009}, these  limitations do not apply to optical lattices and the merging transition was previously observed \cite{tarruell_creating_2012,duca_2014}. In a graphene lattice with inversion symmetry, the band gap is protected by symmetry and doesn't open until the merging of the topological defects \cite{zhu_simulation_2007}. In this case, the merging transition is accompanied by a transition from a semimetal to an insulator and the clear effect on the band structure can be used to detect the transition \cite{tarruell_creating_2012}. In a lattice with broken inversion symmetry as in our case, there is always a band gap. Then the merging transition is purely topological in nature with a dramatic change only in the phase profile. Therefore our method is well suited to map out this phase transition.

The position of the topological defects fits very well to the expectation from band structure calculations (Fig.~\ref{fig:4_mergingTransition}). We find the transition at a critical beam imbalance of $I^{(1)}/I^{(2)}=1.18$ ($I^{(3)}=I^{(2)}$ throughout), which corresponds to a critical imbalance of the next-neighbor tunneling elements of $J_{AB}^{(1)}/J_{AB}^{(2)}=2.05$. This critical ratio of tunneling elements is close to the prediction of $J_{AB}^{(1)}/J_{AB}^{(2)}=2.0$, which is valid if the next-nearest-neighbour tunneling can be ignored. This can be seen from a generalization of the argument in Ref.~\cite{hasegawa_zero_2006,zhu_simulation_2007,pereira_tight-binding_2009} to a lattice with sublattice offset. Our measurements show that the topological defects are robust under the change of lattice parameters and can only be destroyed by annihilation of the two defects of opposite winding.

	Due to its versatility our method could be extended in several directions. It should be applicable to Floquet systems with a large Floquet driving frequency, such that a stroboscopic but still time-resolved measurement is possible. As our method does not couple to the spin degree of freedom, it could be used to detect spin-dependent Dirac points or non-Abelian gauge fields \cite{osterloh_cold_2005,ruseckas_non-abelian_2005} by combining it with Stern-Gerlach separation. Our method is also suited to detect higher winding numbers of the pseudospin textures, which are predicted for driven graphene \cite{sentef_theory_2015} and bilayer graphene \cite{min_chiral_2008,kumar_flat_2013}. Finally, using two driving frequencies, an extension to lattices with a three-atomic basis such as the Lieb lattice \cite{taie_coherent_2015} or generally to multiband systems is conceivable. As our method works with a completely filled band instead of accelerated wave packets, it might be a suitable starting point for the characterization of the topology in interacting systems.

We thank Jean-Noel Fuchs and Lih-King Lim for fruitful discussions. We acknowledge funding from the Deutsche Forschungsgemeinschaft through the excellence cluster The Hamburg Centre for Ultrafast Imaging - Structure, Dynamics and Control of Matter at the Atomic Scale, the GrK 1355 and the SFB 925. B. R. acknowledges financial support from the European Commission (Marie Curie Fellowship).


%
\newpage

\renewcommand{\theequation}{S\arabic{equation}}
\renewcommand{\thefigure}{S\arabic{figure}}
\renewcommand{\bibnumfmt}[1]{[S#1]}
\renewcommand{\citenumfont}[1]{S#1}
\setcounter{figure}{0}  
\setcounter{equation}{0}  
		\section{Initial Hamiltonian and steady state}\label{app:initialHamiltonian}
			We use the fermionic tight-binding Hamiltonian for one non-interacting spin species with a potential offset between the A and the B sites
				\begin{align*}
					H_{\rm init}&=H_J+H_\Delta\\
					H_J&=-J_{AB}\sum_{\ave{{\ve n}x,{\ve m}y}_{\rm n}} c_{{\ve n},x}^\dag c_{{\ve m},y}+ J_{AA} \sum_{\ave{{\ve n}A,{\ve m}A}_{\rm nn}}  c_{{\ve n},A}^\dag c_{{\ve m},A}\\
					H_\Delta&=\Delta_{AB}/2\, \sum_{{\ve n}} c_{{\ve n},A}^\dag c_{{\ve n},A} - c_{{\ve n},B}^\dag c_{{\ve n},B}
				\end{align*}
			where $c_{{\ve n},x}$ annihilates a particle in the ${\mbf n}$-th unit cell at the sublattice site $x=A,B$ and obeys the anti-commutator $\{c_{{\ve n},x}^\dag,c_{{\ve m},y}\}=\delta_{{\ve n}{\ve m}}\delta_{xy}$ for $x=A,B$. The Hamiltonian has two contributions: the hopping part $H_J$ with nearest- and next-nearest-neighbour hopping $J_{AB}$ and $J_{AA}$ and the potential-offset part $H_\Delta$ with $\Delta_{AB}$ being the energy offset between the A and B sublattice sites. The next-nearest-neighbour hopping between the B sublattice sites $J_{BB}$ is negligible for $\Delta_{AB}\gg J_{AB}$. Finally, $\ave{{\ve n}x,{\ve m}y}_{n}$ denotes the sum over nearest neighbours and $\ave{{\ve n}A,{\ve m}A}_{\rm nn}$ the sum over next-nearest neighbours.

\begin{figure}[tb]
				\centering
			    \includegraphics[width=\linewidth]{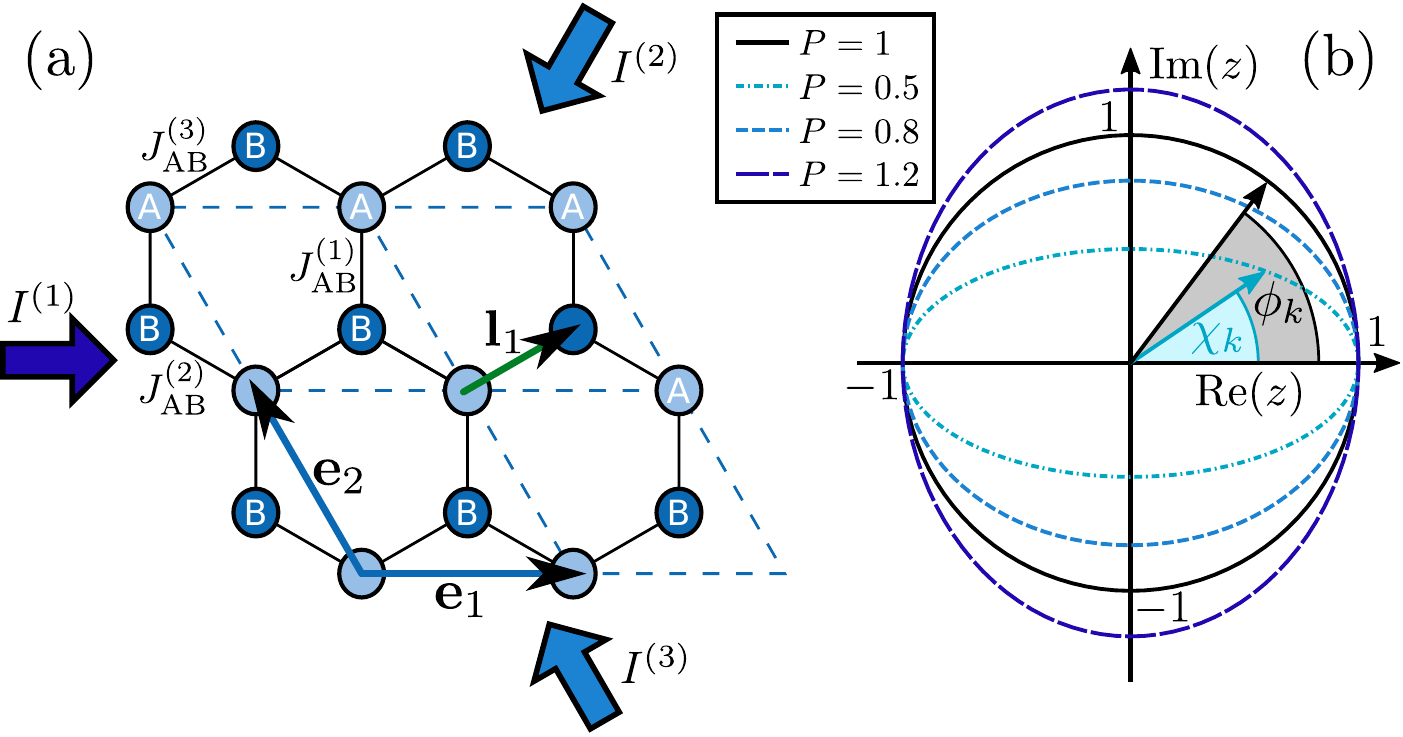}
			    \caption{(a) Sketch of the hexagonal lattice with unit vectors $\mbf e_1=(a,0)^T$ and $\mbf e_2$ and lattice constant $a$. Solid black lines show the hexagonal lattice of Wigner-Seitz unit cells. The dashed blue parallelograms show an equivalent lattice spanned by $\mbf e_1$ and $\mbf e_2$. Both lattices contain two sites, $A$ and $B$, per unit cell, which are connected by the vector $\mbf l_1$. The lattice is formed by the interference of three laser beams. Unequal intensities $I^{(1)}$, $I^{(2)}$, $I^{(3)}$ of these beams lead to unequal nearest-neighbour hopping strengths $J_{\rm AB}^{(1)}$, $J_{\rm AB}^{(2)}$ and $J_{\rm AB}^{(3)}$. 
			    (b) Illustration of the relation of the measured phase $\chi_{\ve k}$ and the azimuthal phase $\phi_{\ve k}$. We plot the real and imaginary part of $z=\cos(\phi_{\ve k})+iP_{\ve k} \sin(\phi_{\ve k})$ for several values of $P_{\ve k}$ as indicated in the legend. Each angle $\phi_{\ve k}$ is represented by the complex phase of the corresponding point on the $P=1$ circle as shown by the grey shaded region in the figure. The related angle $\chi_{\ve k}$ is then obtained by choosing the point on the circle with the appropriate $P$ value that has the same value of ${\rm Re}(z)$.
			    We see that for each $\phi_{\ve k}$ the vector $\cos(\phi_{\ve k})+i P_{\ve k}\sin(\phi_{\ve k})$ remains in the same quadrant of the complex plane for all $P_{\ve k}$ and therefore the phase winding of $\phi_{\ve k}$ is preserved in $\chi_{\ve k}$.
			    }
			    \label{fig:lattice}
			\end{figure}

			In the remaining part of this chapter we present the diagonalization of the Hamiltonian. In a first step we block diagonalize the Hamiltonian by introducing the quasi-momentum operators $c_{{\ve k},x}=\frac{1}{\sqrt M}\sum_{n} e^{-i\mbf k (\mbf n+\delta_x)} c_{{\ve n},x}$. Here, $\delta_A=0$ and $\delta_B=\mbf l_1$, $\mbf l_1$ is defined in Fig.~\ref{fig:lattice}(a) and $M$ is the number of unit cells. With these definitions the Hamiltonian in momentum space is
			\begin{align*}
				H_{\rm init}&=\sum_{\ve k}\bbm c_{{\ve k},A}^\dag & c_{{\ve k},B}^\dag \ebm \bbm \Delta_{AB}/2+2 J_{AA} g_{\ve k} & - J_{AB} f_{\ve k}^* \\ - J_{AB} f_{\ve k} & -\Delta_{AB}/2 \ebm \bbm c_{{\ve k},A} \\ c_{{\ve k},B}\ebm\; , \numberthis\label{initialHamiltonian}
				\intertext{where}
				f_{\ve k}&=e^{-i\mbf k \mbf l_1} \vb{1+ e^{i\mbf k \mbf e_1}+e^{i\mbf k (\mbf e_1 + \mbf e_2)}}\quad \com\\
				g_{\ve k}&=\vb{\cos\vb{\mbf k \mbf e_1}+ \cos\vb{\mbf k \mbf e_2}+\cos\vb{\mbf k (\mbf e_1 +\mbf e_2)}}
			\end{align*}
			and the basis vectors $\mbf e_1$ and $\mbf e_2$ are defined in Fig.~\ref{fig:lattice}a. The Eigenvalues of $H_{\rm init}$ are $E_{{\ve k}\pm}=J_{AA} g_{\ve k} \pm \epsilon_{\ve k}$ and the Eigenvectors are given in Eq.~2 in the main text, where
			\begin{align*}
				\theta_{\ve k} &=2\, {\rm arccos}\vsb{ \frac{|J_{AB} f_{\ve k}|}{\sqrt{(\epsilon_{\ve k} -\Delta_{AB}/2 -J_{AA} g_{\ve k})^2+|J_{AB} f_{\ve k}|^2}}}\com\\
				\epsilon_{\ve k}&=\sqrt{(\Delta_{AB}/2+J_{AA} g_{\ve k})^2+|J_{AB} f_{\ve k}|^2}
			\end{align*}
			and $\phi_{\ve k}$ is the complex phase of $f_{\ve k}$. Using these defintions and neglecting a momentum-dependent energy shift the Hamiltonian in momentum space can be rewritten in the form of Eq.~1 in the main text.

		\section{Time of flight image for lattices with two lattice sites per unit cell}\label{tofTwoLatticeSites}
			For a time-of-flight (ToF) image the atoms are released from the trap and expand freely. For sufficiently long expansion times, such that the cloud is much larger than the initial system size, the initial spatial position of the atoms is irrelevant. Instead, the time-of-flight image is determined by the initial momentum distribution. For a lattice with only one site per unit cell there is an effective mapping between real and momentum space. In addition to this effect the atoms of the A- and B-sublattice interfere for a lattice with two sites per unit cell. In fact, we get for the density measured after ToF
			\begin{align}
				\rho_{\rm TOF}(\mbf r, \tau)&\sim n_{\rm TOF}(\mbf k=\frac{m \mbf r}{\tau \hbar})\\
				&=n_{{\ve k},A}+n_{{\ve k},B} +(\ave{c_{{\ve k},A}^\dag c_{{\ve k},B}}+c.c.)\dt \label{eq:tof}
			\end{align}
			where $n_{{\ve k},x}=\ave{c_{{\ve k},x}^\dag c_{{\ve k},x}}$ and $x=A,B.$ 

		\section{Driven Hamiltonian}\label{app:drivenHamiltonian}
			In addition to the initial Hamiltonian we now apply a periodic driving which can be represented by making all tight binding parameters time dependent
			\begin{align*}
				J_{AB}&\rightarrow J_{AB}+ J_{AB}^{(d)} \sin(\omega t)\\
				J_{AA}&\rightarrow J_{AA}+J_{AA}^{(d)} \sin(\omega t)\\
				\Delta_{AB}&\rightarrow \Delta_{AB}+\Delta_{AB}^{(d)} \sin(\omega t)\dt
			\end{align*}
			For a modulation of the laser beam intensities $\Delta_{AB}$ is modulated out of phase with respect to $J_{AB}$ and $J_{AA}$ ($\Delta_{AB}^{(d)}<0$).
			The full Hamiltonian in the basis that diagonalizes the initial Hamiltonian can then be written as
			\begin{align}
				H&= \bbm c_{{\ve k},+}^\dag & c_{{\ve k},-}^\dag \ebm (H_{\rm init }+ H_{\rm d}) \bbm c_{{\ve k},+} \\ c_{{\ve k},-}\ebm \label{fullHamiltonian}\\
				H_{\rm init}&= \bbm E_{{\ve k},+} & 0 \\ 0 & E_{{\ve k},-} \ebm \\
				H_{\rm d}&=\sin(\omega t)  \bbm E_{{\ve k},+}^{(\rm d)} & E_{\ve k}^{(\rm d)} \\ E_{\ve k}^{(\rm d)} & E_{{\ve k},-}^{(\rm d)} \ebm \label{app:eq:fullHamiltonianDr} \dt
			\end{align}
			For the Hamiltonian from Eq.~\ref{initialHamiltonian} we obtain
			\begin{align*}
				E_{{\ve k},\pm}^{(\rm d)}&=J_{AA}^{(d)} g_{\ve k} \pm \frac{J_{AB}J_{AB}^{(d)} |f_{\ve k}|^2+(\frac{\Delta_{AB}}2+J_{AA}g_{\ve k}) (\frac{\Delta_{AB}^{(d)}}2+J_{AA}^{(d)} g_{\ve k}) }{\epsilon_{\ve k}}\\
				E_{\ve k}^{(\rm d)}&=\frac{J_{AB}|f_{\ve k}|(\frac{\Delta_{AB}^{(d)}}2+J_{AA}^{(d)} g_{\ve k})-J_{AB}^{(d)} |f_{\ve k}|(\frac{\Delta_{AB}}2+J_{AA}g_{\ve k})}{\epsilon_{\ve k}}\dt
			\end{align*}

		\section{Time of flight expression in perturbation theory}\label{app:perturbativeTOF}
			We assume that the corrections to the initial Hamiltonian are small and apply time-dependent perturbation theory in the Heisenberg picture. For an uncorrelated initial state, i.e. $\ave{c_{{\ve k},+}^\dag c_{{\ve k},-}}=0$, we obtain to first order in the perturbation Hamiltonian for the expectation values of the time evolved operators
			\begin{align*}
				\ave{c_{{\ve k},+}^\dag(t) c_{{\ve k},+}(t)}&\approx n_{{\ve k},+}\\
				\ave{c_{{\ve k},-}^\dag(t) c_{{\ve k},-}(t)}&\approx n_{{\ve k},-}\\
				\ave{c_{{\ve k},+}^\dag(t) c_{{\ve k},-}(t)}&\approx \frac{E_{\ve k}^{(\rm d)}}{4\epsilon_{\ve k}^2-\omega^2} \times \\
				\Big[2\epsilon_{\ve k} \sin(\omega t)&-i\omega \cos(\omega t) + i\omega e^{2i\epsilon_{\ve k} t}\Big]
				(n_{{\ve k},+}-n_{{\ve k},-})\dt
			\end{align*}
			We now replace the second-quantization operators in Eqs.~4 in the main text by the corresponding time-dependent operators and insert the results from perturbation theory. A short calculation shows that
			\begin{align*}
				n_{\rm TOF}(\mbf k,t)&=n_{{\rm eq},{\ve k}}-\delta n_{\ve k}\sin(\omega t +\chi_{\ve k}) - \delta \tilde n_{\ve k} \sin(2\epsilon_{\ve k} t +\xi_{\ve k})\com
			\end{align*}
			where
			\begin{align*}
				\chi_{\ve k}&={\rm Arg}\vsb{\cos(\phi_{\ve k}) + i P_{\ve k} \sin(\phi_{\ve k})}+{\rm Arg}\vsb{D_{\ve k}}\\
				n_{{\rm eq},{\ve k}}&=n_{{\ve k},+}+n_{{\ve k},-}+\sin(\theta_{\ve k}) \cos(\phi_{\ve k}) (n_{{\ve k},+}-n_{{\ve k},-})\\
				\delta n_{\ve k}&=\left|D_{\ve k}\sqrt{4\epsilon_{\ve k}^2 \cos^2(\theta_{\ve k})  \cos^2(\phi_{\ve k}) + \omega^2 \sin^2(\phi_{\ve k}) }\right| \\
				P_{\ve k}&=\frac{\omega}{2\epsilon_{\ve k} \cos(\theta_{\ve k})}\\
				D_{\ve k}&=\frac{2E_{\ve k}^{(\rm d)} }{4\epsilon_{\ve k}^2-\omega^2} (n_{{\ve k},-}-n_{{\ve k},+}) \\
				\delta \tilde n_{\ve k}&=|F_{\ve k}| \\
				\xi_{\ve k}&= {\rm Arg}[F_{\ve k}]\\
				F_{\ve k}&=- \omega D_{\ve k} \vsb{\cos(\theta_{\ve k}) \cos(\phi_{\ve k}) +i \sin(\phi_{\ve k}) }\dt
			\end{align*}
			The above equations are valid for an arbitrary Hamiltonian with two lattice sites that is diagonalized by Eq.~2 in the main text.	For any such Hamiltonian the measured phase $\chi_k$ is closely related to the azimuthal phase $\phi_k$ and in particular the two phases have the same winding number (see Fig.~\ref{fig:lattice}(b)).
			
The method is very general, however, a requirement is a finite offset $\Delta_{AB}$, i.e. a broken inversion symmetry. Without this offset, one has $J_{BB}=J_{AA}$. Repeating the above calculation including $J_{BB}$ then yields $E_{\ve k}^{\rm (d)}=0$ and therefore a vanishing oscillation amplitude $\delta n_{\ve k}=0$. The physical origin is that the two bands are not coupled via amplitude modulation in this case.
	
			\begin{figure}[tb]
				\centering
			    \includegraphics[width=\linewidth]{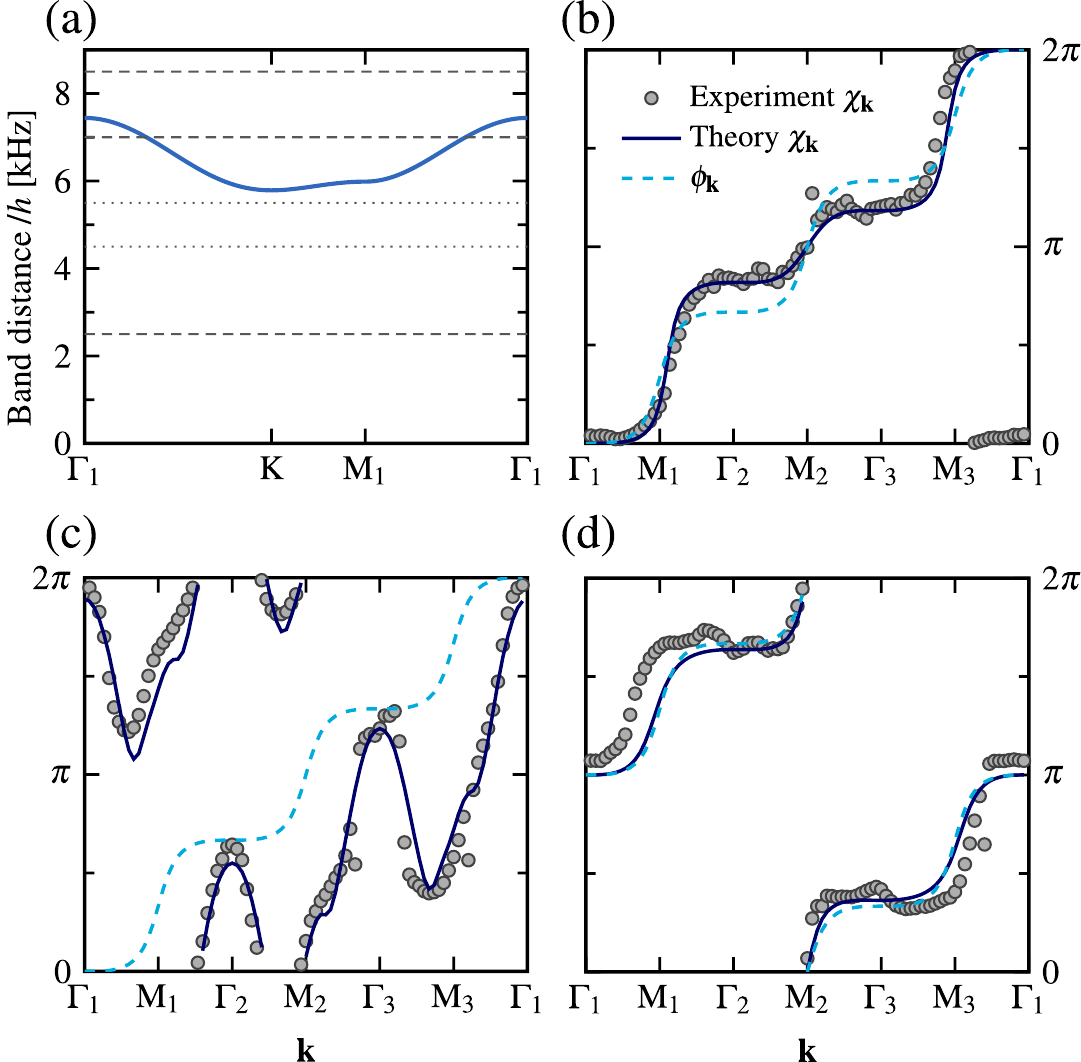}
			    \caption{Comparison of the different driving regimes. (a) Band distance for the used parameters along a high symmetry path in momentum space. Dotted lines indicate the near-resonant red-detuned driving frequencies used for the data in Fig.~3 and Fig.~4 in the main text. Dashed lines indicate exemplary driving frequencies in different regimes as shown in (b)-(d). (b)-(d) Comparison of the experimental phase $\chi_{\ve k}$ (grey points), a theory prediction for $\chi_{\ve k}$ (solid line) and the azimuthal phase $\phi_{\ve k}$ (dashed line) along a high symmetry path in momentum space for different driving regimes. Except for the driving frequency the parameters are the same as in Fig.~3. (b) Red-detuned driving regime, but with larger detuning than in the main text ($\omega= 2\pi\cdot 2500$ Hz). The experimental data follows the prediction of perturbation theory, which however has a slight deformation from the azimuthal phase (compare Fig.~\ref{fig:lattice}b). (c) resonant driving regime ($\omega= 2\pi\cdot 7000$ Hz).  As perturbation theory is not valid on resonance, the theoretical prediction for $\chi_{\ve k}$ is obtained by numerically computing the time of flight density as given by Eq.~\ref{eq:tof} and then fitting a sinusoidal oscillation to the obtained signal. The phase has a widely oscillating shape, which arises from the $\pi$ phase jumps that are washed out due to the fitting procedure. This regime is not practical to infer information on the azimuthal phase. (d) Blue-detuned driving regime ($\omega= 2\pi\cdot 8500$ Hz). There is an overall phase shift of $\pi$. For near-resonant blue detuning, the deformation between $\chi_{\ve k}$ and $\phi_{\ve k}$ is negligible, but there is a deformation in the data, which we attribute to the coupling to higher bands, which is not included in the theory.}
			    \label{fig:drivingRegimes}
			\end{figure}%
			
		\section{Discussion of the different driving regimes}\label{app:drivingRegimes}

			In the main text we have only considered red-detuned driving. In this case the term ${\rm Arg}\vsb{D_{\ve k}}$ in the expression for $\chi_{\ve k}$ vanishes and is therefore omitted in Eq.~6 in the main text. In general the expression for the density after ToF has an additional term oscillating with the band gap $2\epsilon_{\ve k}$. For red-detuned driving, its amplitude is smaller than the term oscillating with the driving frequency. It has therefore been omitted in Eq.~5 in the main text. 

			Our method to measure the azimuthal phase can be applied for any driving frequency. It is, however, unpractical in the resonant regime due to the sign changes of $D_{\ve k}$ which lead to $\pi$ phase jumps in $\chi_{\ve k}$. Furthermore perturbation theory is not valid in that regime. A blue-detuned shaking frequency is possible and the same formulas apply, but the second oscillation with frequency $2\epsilon_{\ve k}$ is not negligible. 

Experimental data and calculations in the different driving regimes are shown and discussed in Fig.~\ref{fig:drivingRegimes}. This comparison confirms that the near-resonant red-detuned regime is the most suitable regime, which is why the discussion is restricted to this regime in the main text. The comparison of data and theory in Fig.~\ref{fig:drivingRegimes} demonstrates that the observed dynamics is well understood in all driving regimes.	

		\section{Data analysis}\label{app:dataAnalysis}
\begin{figure}[!h]
				\centering
			    \includegraphics[width=\linewidth]{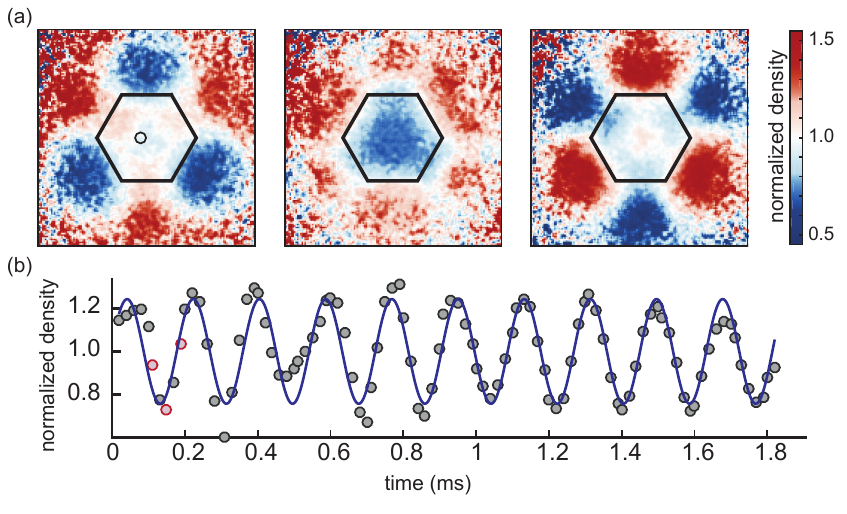}
			    \caption{Experimental data showing the momentum-dependent oscillation with the driving frequency. (a) Normalized density distributions after time-of-flight expansion after three different modulation times. The hexagon marks the first Brillouin zone. The parameters are as in Fig. 3 in the main text. (b) Exemplary density oscillation for a single pixel (indicated by the black circle in (a)) together with the sinusoidal fit (solid line), from which the phase is extracted. The three colored data points indicate the time steps shown in (a). A slight beating is visible, which stems from the second term of the perturbation theory oscillating with the band gap. We found that applying a fit including a second variable frequency yields indiscernible results for the phase $\chi_{\ve k}$. The amplitude of the this second term becomes larger for blue-detuned driving.}
			    \label{fig:dataAnalysis}
			\end{figure}

The data is obtained as absorption images after 21 ms of time-of-flight expansion after different modulation times. We apply a spatial smoothing with a Gaussian filter of 3 pixel width (small compared to the length of the reciprocal lattice vector of 58 pixels) and a temporal smoothing taking a floating average including 3 time steps (time steps are chosen as one tenth of the driving period). Subsequently the average density for each pixel is normalized to one. Exemplary normalized density distributions are shown in Fig.~\ref{fig:dataAnalysis}(a). We perfom a pixelwise sinusoidal fit with variable amplitude, offset and phase. The frequency is fixed to the driving frequency (Fig.~\ref{fig:dataAnalysis}(b)).

\end{document}